# Signatures of chiral superconductivity in chiral molecule intercalated tantalum disulfide

Zhong Wan[1,6], Gang Qiu[2,6], Huaying Ren[1], Qi Qian[1], Dong Xu[3], Jingyuan Zhou[1], Jingxuan Zhou[3], Boxuan Zhou[3], Laiyuan Wang[1], Yu Huang[3,4], Kang L. Wang[2,4,5], Xiangfeng Duan[1,4*]

[1]Department of Chemistry and Biochemistry, University of California Los Angeles, Los Angeles, CA, USA. [2]Department of Electrical and Computer Engineering, University of California Los Angeles, Los Angeles, CA, USA. [3]Department of Materials Science and Engineering, University of California Los Angeles, Los Angeles, CA, USA. [4]California NanoSystems Institute, University of California, Los Angeles, Los Angeles, CA, USA. [5]Department of Physics and Astronomy, University of California, Los Angeles, Los Angeles, CA, USA.
[6]These authors contributed equally: Zhong Wan, Gang Qiu.
*Correspondence: e-mail: xduan@chem.ucla.edu

**Chiral superconductors, a unique class of unconventional superconductors in which the complex superconducting order parameter winds clockwise or counter-clockwise in the momentum space[1], represent a topologically non-trivial system with direct implications for topological quantum computing[2,3]. Intrinsic chiral superconductors are extremely rare, with only a few arguable examples including heavy fermion metals (UTe$_2$, UPt$_3$)[4] and perovskite superconductor Sr$_2$RuO$_4$ (ref. [5]). It has been suggested that chiral superconductivity may exist in non-centrosymmetric superconductors[6], although such non-centrosymmetry is uncommon in conventional solid-state lattices. Chiral molecules with neither mirror nor inversion symmetry have been widely investigated[7], in which the spin degeneracy may be lifted by the molecular chirality[8,9]. Thus, we hypothesize that a combination of superconductivity with chiral molecules may lead to a spin-polarized ground state for realizing chiral superconductivity. Herein we report the first investigation of unconventional superconductivity in chiral molecule intercalated tantalum disulfide (TaS$_2$) hybrid superlattices and reveal key experimental signatures of chiral superconductivity. Little-Parks measurements demonstrate a robust and reproducible half-flux quantum phase shift in both left- and right-handed chiral molecule intercalated TaS$_2$, which is absent in pristine TaS$_2$ or achiral molecule intercalated TaS$_2$, highlighting the essential role of molecular**

**chirality in inducing unconventional superconductivity. The robust half-flux quantum phase shift demonstrates unconventional superconductivity and constitutes strong evidence supporting a chiral superconducting ordering parameter. Critical current measurements at lower temperature reveal a peculiar asymmetric phase shift under opposite supercurrent, with a relative phase difference approaching the unity of $\pi$ at below 0.5 K, further supporting topologically non-trivial superconductivity. Our study signifies the potential of hybrid superlattices with intriguing coupling between the crystalline atomic layers and the self-assembled molecular layers. It could chart a versatile path to artificial quantum materials by combining a vast library of layered crystals of rich physical properties with the nearly infinite choices of molecules of designable structural motifs and functional groups[10].**

Superconductivity is one of the most exciting experimental discoveries in modern condensed matter physics framework and holds substantial potential in diverse technologies. It is generally understood that superconductivity is originated from the condensation of two electrons into Cooper pairs with a certain pairing symmetry[11]. In this regard, superconductors can be generally classified by pairing symmetry, with most known superconductors having symmetric pairing states (e.g., *s*-wave superconductors), which have been predominantly studied over the past century[12]. The recent advances in topological classifications of quantum many-body systems have evoked the re-examination of the symmetry-broken pairing states in superconductivity. Theoretically, such a symmetry-broken mechanism in superconductivity can lead to topologically non-trivial superconductivity, which is closely tied to exotic non-Abelian quasiparticle excitations that can have direct implications for fault-tolerant topological quantum computing[2,3]. As one of the topologically non-trivial systems, chiral superconductivity with unique time-reversal-symmetry-broken properties has drawn much recent attention[1]. **Intrinsic chiral superconductors are extremely rare in nature, with only a few arguable examples including heavy fermion metals ($UTe_2$ and $UPt_3$) and perovskite superconductor $Sr_2RuO_4$** (ref.[4,5]). Theoretical studies suggested that chiral superconductivity could be induced by proximitizing topological insulators with s-wave superconductors in the presence of spin-orbit coupling and in-plane magnetic field[13,14]. However, this intricate setting poses complexities in material preparation and device fabrication, leading to ambiguity in validating chiral superconductivity.

It has been predicted that chiral superconductivity can also exist in non-centrosymmetric superconductors[6]. Although such non-centrosymmetry is uncommon in solid-state lattices, it is a general and widely explored property of chiral molecules with neither mirror nor inversion symmetry[7]. It has been recently suggested and demonstrated that chiral molecules may possess a unique chiral-induced spin selectivity (CISS) effect, in which the spin degeneracy can be lifted depending on the molecular chirality[8,15,16]. Considerable efforts have been made to bring CISS into solid-state systems, for example, by coupling self-assembled chiral molecular monolayers with metals or semiconductors to generate spin selectivity[17–19]. However, the systems explored to date generally suffer from high inhomogeneity and poorly controllable disorder potential that renders the system into a strongly diffusive regime, where quantum phase coherence length can hardly extend into a mesoscopically measurable regime[20]. Thus, the incorporation of molecular chirality in superconducting systems has been a largely overlooked arena[21].

We have recently developed a new class of chiral molecular intercalation superlattices (CMIS), in which selected chiral molecules are intercalated into two-dimensional atomic crystals (2DACs) to form an ordered superlattice structure consisting of alternating layers of crystalline atomic layers and self-assembled molecular layers. Within the CMIS, the crystalline atomic layers provide a natural template to guide the self-assembly of the intercalated molecular species into highly-ordered molecular layers sandwiched between the crystalline atomic layers, and simultaneously function as a protective layer and the electrically contacting layer for robustly integrating the molecular assembly into solid-state devices. Our studies have shown that such CMIS exhibits a robust CISS effect with a spin polarization ratio exceeding 60%, which is much stronger than the intrinsic spin-orbital coupling expected in non-centrosymmetric molecules[9] and suggest possible projection of molecular chirality into the crystalline atomic layers. **It is conceivable that a combination of superconductivity with such chiral molecule induced spin selectivity may lead to a spin-polarized ground state for realizing spin-triplet Cooper pairing and chiral superconductivity.**

Herein, we report the first investigation of unconventional superconductivity in chiral molecule intercalated 2H-phase tantalum disulfide (2H-TaS$_2$) and reveal key experimental signatures in this unique class of hybrid superlattices. We observed a half-flux quantum phase shift in Little-Parks measurements, providing strong evidence of a chiral superconducting ordering parameter. Such a half-flux quantum phase shift is robust and reproducible in opposite chiral molecule intercalated TaS$_2$ devices with different cool-down procedures, but absent in pristine 2H-TaS$_2$ or achiral molecule intercalated 2H-TaS$_2$, highlighting the essential role of molecular chirality in realizing such unconventional superconducting properties. Critical current measurements reveal an asymmetric phase shift under opposite DC current bias, with the relative phase difference approaching π at the lowest temperature, further supporting the emergence of topologically non-trivial superconductivity.

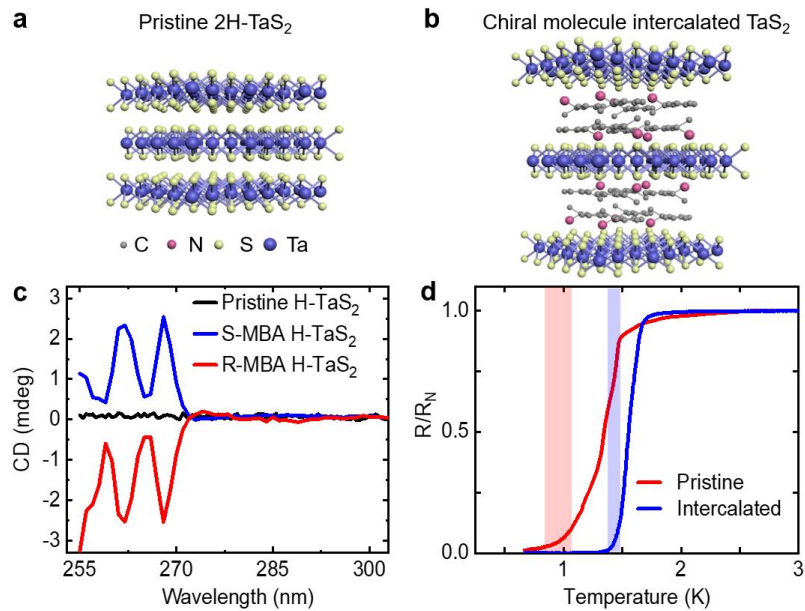

**Fig. 1| The intercalation of chiral molecules into a layered superconductor 2H-TaS$_2$. a, b**, Schematic drawings of pristine 2H-TaS$_2$ (**a**) and chiral molecule-intercalated 2H-TaS$_2$ (**b**). **c**, Circular Dichroism (CD) spectra of right-handed methylbenzylamine (R-MBA) and left-handed methylbenzylamine (S-MBA) chiral molecule intercalated and pristine 2H-TaS$_2$. **d**, Normalized resistance of pristine and R-MBA intercalated 2H-TaS$_2$ as a function of temperature, highlighting the enhancement of $T_C$ upon molecular intercalation in 2H-TaS$_2$. Shaded regions in the superconducting transition tail correspond to the temperature window where the Little-Parks oscillation features are observable.

The R- and S-methylbenzylamine (R- and S-MBA) intercalated 2H-TaS$_2$ superlattices were synthesized using our recently reported chemical intercalation approach (Fig. 1a, b)[9]. The layered 2H-TaS$_2$ represents a rich family of 2DAC hosting systems with intriguing superconductivity and charge density wave phase transition that are highly sensitive to the structural, charge or spin modulation by intercalated molecules[22–26]. The X-ray diffraction patterns of the resulting intercalation superlattices exhibit sharp diffraction peaks with a notable expansion of interlayer spacing from 5.8 Å in pristine 2H-TaS$_2$ to 11.7 Å in hybrid superlattices (Extended Data Fig. 1), suggesting the formation of a highly ordered superlattice structure. Our circular dichroism (CD) spectroscopy studies reveal distinct CD absorption features in the wavelength range of 255–275 nm (Fig. 1c), indicating chirality-dependent preferential absorption of different circularly polarized light and confirming the successful incorporation of the molecular chirality into the hybrid superlattices.

We further conducted temperature-dependent resistivity measurements to explore the superconducting behavior in the pristine 2H-TaS$_2$ and the chiral molecule intercalated 2H-TaS$_2$. Fig. 1d presents the resistance versus temperature curves for representative pristine (red) and intercalated (blue) 2H-TaS$_2$ devices. The pristine 2H-TaS$_2$ device exhibits a superconducting critical temperature $T_C$ (defined by 10% of normal resistance states R$_N$) of 1.1 K, which is in good agreement with previously reported $T_C$ in 2H-TaS$_2$ (ref.[27]). In comparison, the $T_C$ is enhanced to ~ 1.5 K in the intercalated device. This is consistent with the previous suggestion that molecular intercalation can effectively suppress the interlayer interactions and strengthen electron-phonon coupling to boost $T_C$ (ref. [28,29]). A similar enhancement of the $T_C$ was also observed in other molecular intercalated 2DACs[27,30]. Thus, the enhancement of superconductivity confirms the successful molecular intercalation in 2H-TaS$_2$ as well.

To probe the key signatures of the unconventional pairing symmetry in the chiral superconductivity, a phase-sensitive measurement is essential[31]. The evaluation of the Little-Parks effect in superconducting rings offers such a measure. The Little-Parks effect describes the periodic oscillation of free energy as a function of magnetic flux encircled by a superconducting ring. In a superconductor, the fluxoid $\Phi'$ inside the superconductor has the following expression:

$$\Phi' = \Phi + \frac{4\pi}{c} \oint \lambda^2 \vec{j} d\vec{s}$$

Where $\Phi$ is the applied magnetic flux, $c$ is the speed of light, $\lambda$ is the London penetration depth and $\vec{j}$ is the supercurrent density[12]. The applied magnetic flux can take arbitrary values, while fluxoid $\Phi'$ has to be quantized, with $\Phi' = n\Phi_0$, where $\Phi_0 = \frac{h}{2e}$ is the magnetic flux quantum, $n$ is an integral number, $h$ is the Planck's constant, $e$ is the charge of an electron. The difference between the quantized fluxoid $\Phi'$ and the applied magnetic flux $\Phi$ is compensated by a spontaneous supercurrent $\vec{j}$, which leads to a change of free energy $\Delta f_N$, where $\Delta f_N \propto \vec{j}^{\,2} \propto (\Phi' - \Phi)^2$. This relation results in a periodic oscillation of $\Delta f_N$ with the applied magnetic flux.

At a fixed temperature slightly below $T_C$, where superconductivity starts to percolate the ring, the resistance of the superconductor $R$ is very sensitive to the free energy change, with $R \propto \Delta f_N$. This establishes an experimental approach to correlate the resistance oscillations to the free energy oscillations, providing phase-sensitive information through resistance measurement as a function of external magnetic flux. **For conventional superconductors without pairing asymmetry, the symmetry requirement guarantees a zero fluxoid at zero magnetic field, leading to a minimized free energy change thus a resistance minimum at zero field**[32]**. However, for a chiral superconductor, the pairing anisotropy from chirality is expected to generate additional phase shift, which can be detected through resistance measurement in the Little-Parks devices, and regarded as a key signature for chiral superconductivity**[33–36]**.

To explore such signatures of chiral superconductivity in the chiral molecule intercalated 2H-TaS$_2$, we fabricated sub-micron-sized superconducting rings (Fig. 2a, b) to investigate the phase of superconducting order parameter through the Little-Parks effect (see Method section for details on device fabrication). The encircled ring area $S$ (620 nm × 620 nm) defines the required magnetic field ($\Delta B$) to change one unit of flux quantum by $\Delta B = \frac{\Phi_0}{S} = 53.7\ Oe$. The ring size is specifically designed to satisfy three key criteria[34]: (i) it should be small enough so a flux quantum would require a magnetic field at least one order larger than the zero-field offset (~2 Oe) induced by flux trapped in the superconducting magnet of the measurement system; (ii) it should be large enough to access multiple periods of oscillations within the lower critical field ($H_{c1}$); and (iii) it should be large enough to rule out the normal-state Aharonov-Bohm effect, which can be suppressed if the perimeter is larger than the single electron phase coherence length (typically < 10-20 nm)[37].

The magnetic field-dependent resistance measurement of a representative R-MBA intercalated ring device is shown in Fig. 2c. The magnetic field scan was performed at different temperatures slightly below the transition temperature (see the tail of the R-T curves highlighted in Fig. 1d), as the nano-patterned narrow ring region usually has a slightly lower $T_C$ compared to the neighboring superconducting reservoirs. The overall resistance contour of the chiral-TaS$_2$ superlattice shows a symmetric "U" shape with respect to zero magnetic field (Fig. 2c), indicating no significant magnetic zero-field offset in our measurement system. Notably, the resistance measured between 1.3 K to 1.4 K shows a clear periodic oscillation with the magnetic field, consistent with the expected Little-Parks effect[32].

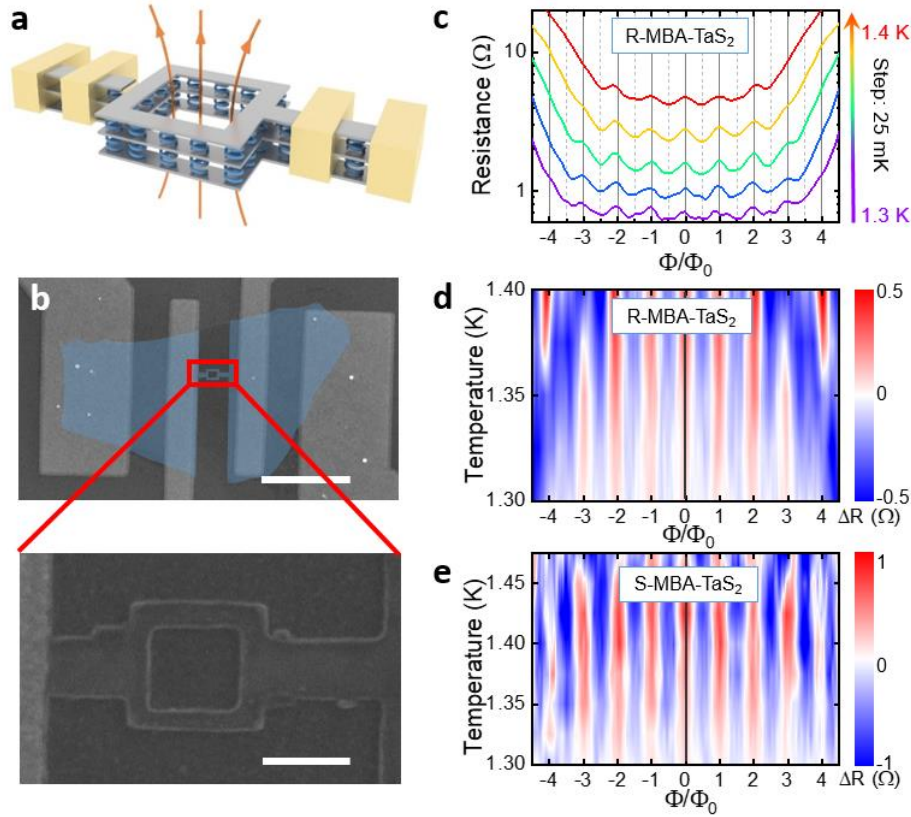

**Fig. 2| Device images and Little-Parks effect with π phase shift. a**, The schematic drawing of a Little-Parks device made of chiral molecule intercalated 2H-TaS$_2$. **b**, Top, a scanning electron microscopy (SEM) image of a Little-Parks device with four-terminal measurement probes. Scale bar: 5 µm. Bottom, a zoomed-in SEM image of the same device. Scale bar: 500 nm. **c**, Four-terminal resistance (logarithmic scale) of an R-MBA intercalated 2H-TaS$_2$ Little-Parks device as a function of magnetic field at different temperatures. **d** and **e**, Color mapping of the Little-Parks oscillation amplitudes of right- (**d**) and left-(**e**) handed chiral molecule intercalated devices as a function of the temperature and the external magnetic flux. Resistance maxima at zero magnetic field are observed, highlighting a π phase shift and signifying unconventional superconductivity.

After subtracting a smooth background, we obtained more pronounced oscillation patterns ($\Delta R$) plotted in a color map in the temperature-external magnetic flux parameter space. A well-defined period of 49.0 Oe (Fig. 2d) can be extracted, which is close to the designed unit of flux quantum (53.7 Oe) defined by the encircled area in the superconducting ring structure. The slight deviation in the oscillation period can be attributed to the finite width of the ring. Notably, **the superconducting ring shows a resistance maximum at zero magnetic field, in contrast to the typically expected/observed resistance minimum at zero field in topologically trivial *s*-wave superconductors**[32,38]. This resistance maximum at zero field corresponds to a half-flux quantum in the ground state, indicating **an intrinsic π phase shift is energetically favorable** in the superconducting ordering parameter of the chiral molecule-intercalated TaS$_2$ superlattices. The oscillations are consistently observed in a narrow temperature window near $T_C$ with maxima/minima located at the same field insensitive to the temperatures, but gradually diminishes at temperature further away from the $T_C$ (Fig. 2c, d).

Furthermore, a similar π phase shift is also observable in the devices intercalated with chiral molecules of opposite chirality (S-MBA) (Fig. 2e, Extended Data Fig. 2). Additionally, we performed similar transport measurements after several thermal cycles, in which chiral-TaS$_2$ superlattice devices were warmed up to above $T_C$ and cooled back down to the base temperature. The Little-Parks devices show consistent oscillation periodicity and π-phase shift after such thermal cycles (Extended Data Fig. 3), indicating that the π-phase shift persists after the vanishing and restoring of the superconducting phase and highlighting the robustness of the observed phenomena.

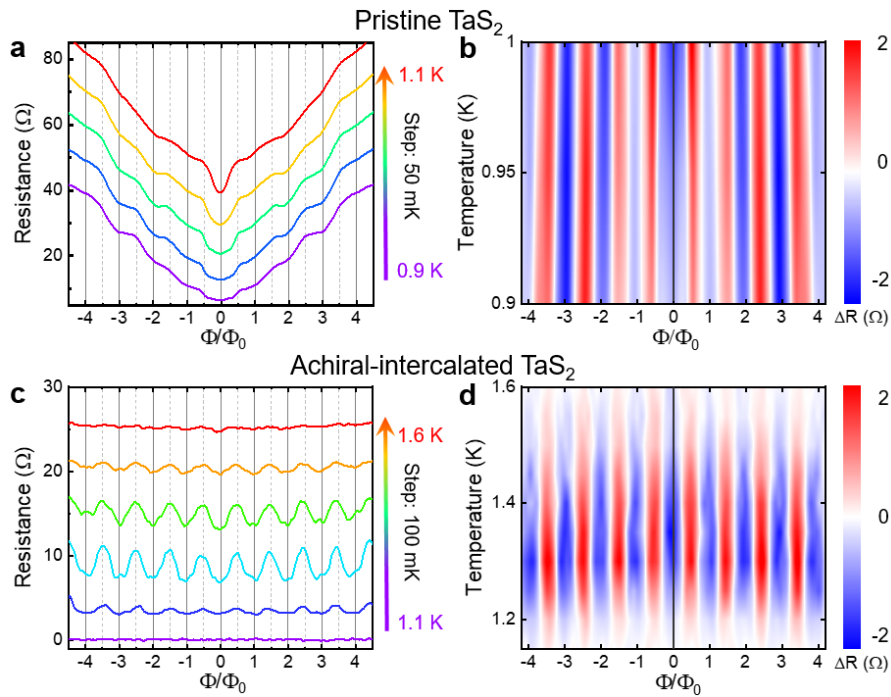

**Fig. 3| Little-Parks oscillations in pristine 2H-TaS$_2$ and achiral molecule intercalated 2H-TaS$_2$. a**, Four-terminal resistance of a pristine 2H-TaS$_2$ device versus external magnetic flux at different temperatures. **b**, Color mapping of Little-Parks oscillation amplitudes of the non-intercalated TaS$_2$ device as a function of external magnetic flux and temperature. **c**, Four-terminal resistance of an achiral molecule (TEAB) intercalated device versus external magnetic flux at different temperatures. **d**, Color mapping of Little-Parks oscillation amplitudes of the achiral molecule intercalated device as a function of external magnetic flux and temperature, highlighting resistance minimum at zero field.

We have also investigated additional control devices fabricated from pristine 2H-TaS$_2$ (without molecular intercalation) and achiral molecule tetraethylammonium bromide (TEAB) intercalated 2H-TaS$_2$ (see Method for details, and Extended Data Fig. 4). In the case of pristine TaS$_2$ devices, the Little-Parks oscillations are also observed near the superconducting transition temperature (Fig. 3a and 3b), with a period of 49.3 Oe, largely consistent with one unit of flux quantum of the defined ring geometry. However, **a resistance minimum is observed at zero field, which is consistent with the previous experiments and the theoretical expectation of the typical *s*-wave superconductivity in the pristine TaS$_2$ materials**[39,40]. Similarly, for the device made of the achiral TEAB intercalated 2H-TaS$_2$ (Fig. 3c and 3d), the Little-Parks oscillations (with a period of 48.5 Oe) **also showed a resistance minimum at zero field**, **which is in stark contrast to chiral molecule intercalated TaS$_2$ showing resistance maximum at zero field.** The absence of a half-flux quantum in achiral molecule intercalated device indicates that the intercalation process itself is not the origin of the topological phase. Multiple devices were fabricated with pristine or TEAB intercalated 2H-TaS$_2$, which all showed resistance minimum at zero field (Extended Data Fig. 5), highlighting the essential role of molecular chirality in inducing unconventional superconductivity and achieving π-phase shifts and resistance maximum at zero field.

When the chiral molecules are intercalated into 2DACs, the chiral property from molecules may be transferred into the crystalline 2D atomic layers, leading to a reconstruction of electronic properties. It has been recently proposed that molecular level chirality, when incorporated into the crystal field, can be projected into its momentum space and thus the superconducting band structures, leading to a chiral superconductivity with topological features[41,42]. From a semiclassical perspective, in a chiral superconductor, the phase of its complex superconducting gap function $\Delta(\mathbf{k})$ winds clockwise or counter-clockwise as $\mathbf{k}$ moves on the fermi surface. For the simplest case of a 2D chiral superconductor, the gap functions can be expressed as a 2D $k_x \pm ik_y$ superconductor[1] (Fig. 4a). Thus, the Bogoliubov-de Gennes (BdG) Hamiltonian $H_{\text{BdG}}(\mathbf{k})$ of this 2D chiral superconductor in the continuum limit can be expressed as[2]:

$$H_{\text{BdG}}(\mathbf{k}) = \begin{pmatrix} \varepsilon(\mathbf{k}) & 2i\Delta(k_x \pm ik_y) \\ -2i\Delta^*(k_x \mp ik_y) & -\varepsilon(\mathbf{k}) \end{pmatrix} \qquad (1)$$

Here $\varepsilon(\mathbf{k})$ is the quasiparticle excitation energy with momentum $\mathbf{k}$, and the off-diagonal terms contain the complex gap function $\mathbf{\Delta} = |\Delta|e^{i\theta_k}$, where $\theta_k = \tan^{-1}(k_x/k_y)$ is the azimuth angle in the momentum space. The choice of the $\pm$ sign depends on the chirality of the chiral superconductors.

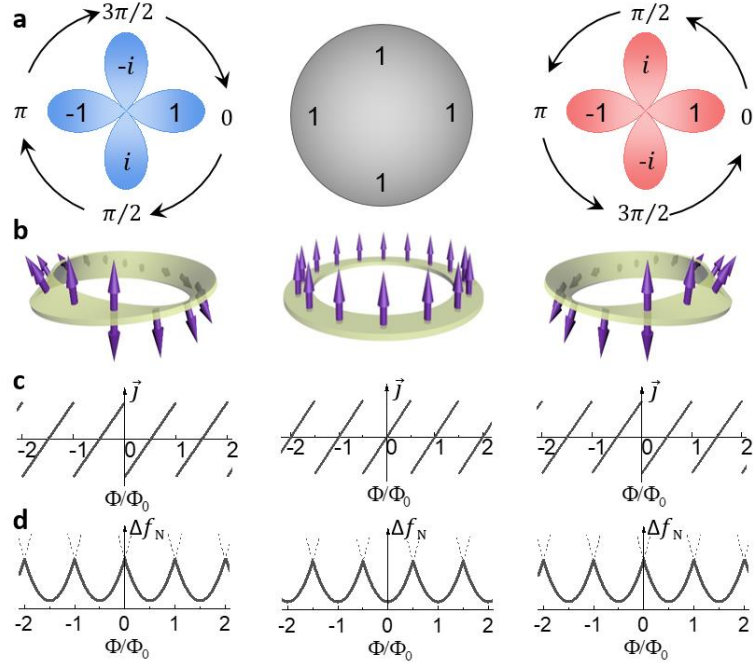

**Fig. 4| Schematic illustrations of chiral superconductivity and phase sensitive measurement. A**, Schematic drawing of the superconducting order parameter in the momentum space, with the left and right columns representing chiral superconductors with opposite chirality, whereas the middle column showing the case for an *s*-wave superconductor. **B**, Schematic drawing of the Berry's phase accumulated along a closed-loop path. **C,** Supercurrent density $\vec{j}$ as a function of external magnetic flux $\Phi$. **d,** The change of free energy $\Delta f_N$ as a function of magnetic flux $\Phi$.

It should be noted that this BdG Hamiltonian shares the same form as that of a massive Dirac system[2]. It has been shown that in a massive Dirac dispersion, for example a gapped bilayer graphene, additional non-zero Berry's phase can occur with $\phi_B = \pi W(1 - \varepsilon/\Delta_S)$ at Dirac points, where $\Delta_S$ is the Dirac mass gap, and $W = \pm 1$ denotes the winding chirality[43] (Fig. 4b). The same topological singularity also holds in the BdG Hamiltonian of a chiral superconductor, thus the corresponding Berry's phase of $\phi_B = \pi W(1 - \varepsilon/\Delta)$ can also be accumulated around the Fermi surface in the chiral superconductors. For low-energy quasiparticle excitations $\varepsilon \to 0$, a Berry's phase $\phi_B = \pi$ is expected along a closed-loop ring structure, which leads to the wave function changing sign with $\varphi(2\pi) = e^{i\phi_B}\varphi(0) = -\varphi(0)$ (ref.[44]) as manifested by the $\pi$-phase shift in the Little-Parks experiments (Fig. 4c, d).

Next, we discuss other possible origins of such π-phase shift from a conventional *s*-wave TaS$_2$ after the chiral intercalation process. It has been well studied that when a superconductor is coupled to a ferromagnet material, the Zeeman energy can induce a finite phase shift. In chiral molecules, it is possible to generate a spin-polarization-induced effective magnetic field[45]. However, such a ferromagnet coupling-induced phase shift is expected to be an arbitrary value ranging from 0 to π, depending on the coupling strength between the superconductor and the ferromagnet. In this case, the system would not have a well-defined phase shift, thus rendering a topologically trivial one[46,47]. In contrast, our Little-Parks measurements show a consistent π-phase shift across multiple CMIS devices, suggesting a non-trivial superconducting order parameter originated from the topological nature of the CMIS.

In addition to the Berry curvature-induced topological phase shift, it is also possible that such phase shift can exist in the domain boundary between the different degenerate superconducting states, which can occur inside superconducting rings made of unconventional superconductors with polycrystalline domains[36,48,49]. Although we cannot completely exclude such a possibility in our devices, this is less likely considering the single crystalline nature of 2H-TaS$_2$ flake used to fabricate our superconductor ring devices. Further, it should be noted the π-phase shift is not expected from a conventional superconductor with *s*-wave pairing symmetry, regardless of the presence of grain boundaries. Indeed, we did not observe π-phase shift in pristine or achiral molecular intercalated 2H-TaS$_2$ fabricated with similar protocols, highlighting that the chiral components play an indispensable role in the observed π-phase shift (or unconventional superconductivity).

In addition to the resistance oscillations in the vicinity of $T_C$, the π-phase shift can also be confirmed by measuring the critical current ($I_C$) of the superconducting ring modulated with external flux below $T_C$. Fig. 5a shows the color mapping of the critical current as a function of magnetic field at *T*=1.1 K. **A minimum of critical current is observed at zero magnetic field, suggesting a minimum superconducting gap without any external magnetic field, which is consistent with the π-phase shift observed in the Little-Parks oscillations shown in Fig. 2. This π-phase shift of the critical current remains unchanged when the polarity of the applied current is reversed at 1.1 K.**

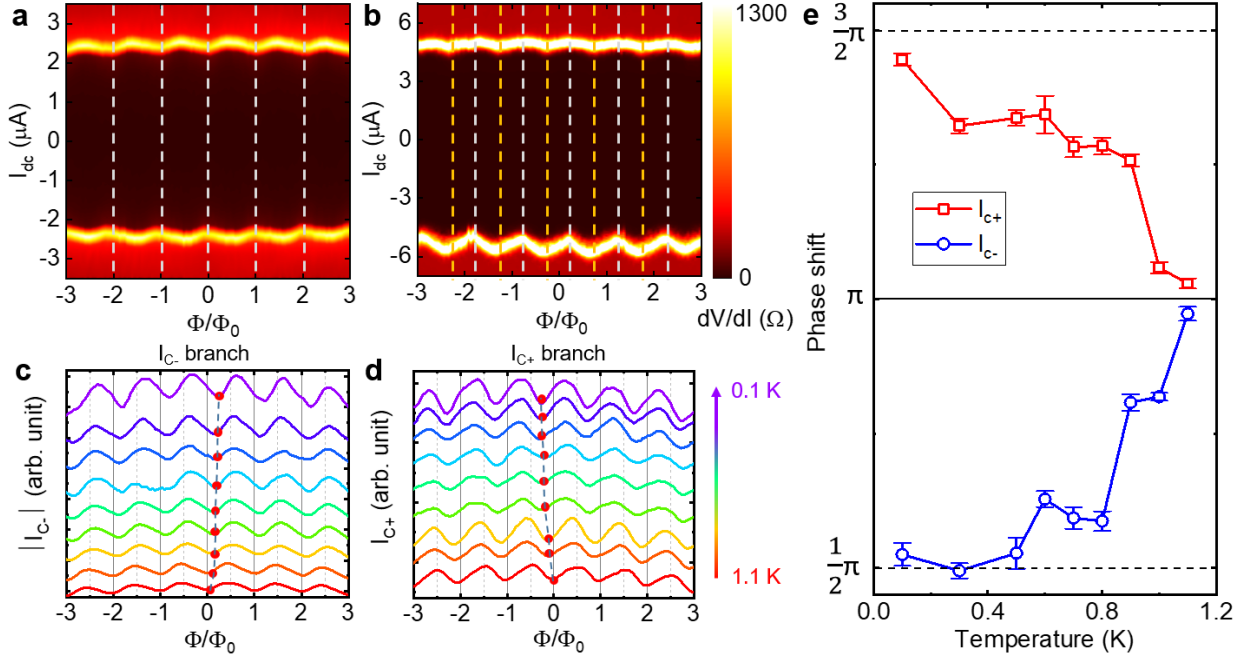

**Fig. 5| Critical current as a function of magnetic field below $T_C$. a** and **b**, Color mapping of the differential resistance of the R-MBA intercalated 2H-TaS$_2$ rings as a function of the temperature and the external magnetic flux, at T=1.1 K (**a**), and T=0.1 K (**b**). **c** and **d**, Critical current of negative bias, $I_{C-}$ (**c**) and positive bias $I_{C+}$ (**d**) as a function of external magnetic flux at different temperatures, showing the $I_{C-}$ and $I_{C+}$ are shifted toward positive and negative magnetic field, respectively, with reducing temperature. **e,** The phase shift of $I_{C-}$ (red) and $I_{C+}$ (magenta) as a function of temperature highlights an asymmetric phase shift approaching $\frac{3}{2}\pi$ and $\frac{1}{2}\pi$ at lower temperature. Error bars: standard deviation from linear fitting as described in Methods.

Further measurements at 0.1 K reveal an intriguing additional phase shift in opposite directions for positive critical current ($I_{C+}$) and negative critical current ($I_{C-}$) (Fig. 5b). The peaks in the critical current in one current direction are now aligned with valleys in the other current direction (dashed lines in Fig. 5b), in contrast to the synchronized oscillations at higher temperatures (dashed lines in Fig. 5a). Such asymmetric phase shifts in opposite directions evolve continuously with decreasing temperature, as highlighted by tracing the minima near zero field in Fig. 5c and 5d. We further extracted and plotted the phase shift in both positive and negative current branches as a function of temperature (Fig. 5e, Extended Data Fig. 6). It is evident that the phase shifts for $I_{C+}$ and $I_{C-}$ gradually divert from π at 1.1 K and increasingly approach $\frac{3}{2}\pi$ and $\frac{1}{2}\pi$ at lower temperatures. While similar asymmetric phase shifts under opposite supercurrent have also been reported in the asymmetric ring geometry (e.g., from intentional design or fabrication non-ideality), the typical phase shifts in such cases usually exhibit a much smaller and arbitrary value (~0.2 π at 0.05 K) [50,51]. Our case is notably different, showing an apparent **saturation** **towards $\frac{3}{2}\pi$ and $\frac{1}{2}\pi$ in phase shift with a net phase difference of nearly a unity of π below 0.5 K, which implies a non-trivial topological origin and may be fundamentally related to its chiral superconductivity nature.** Nonetheless, the exact reason for such asymmetric phase shift remains open for interpretation at this stage and requires further investigation.

Together, we have reported the vital signatures of spin triplet pairing and chiral superconductivity in chiral molecule intercalated 2H-TaS$_2$. We observed a half-flux quantum phase shift in Little-Parks measurements, providing key evidence supporting the chiral superconducting ordering parameter. Such a π phase shift is robust and reproducible in multiple TaS$_2$ devices with both left- and right-handed chiral molecule intercalation and with different cool-down procedures, but absent in pristine 2H-TaS$_2$ and achiral molecule intercalated 2H-TaS$_2$, highlighting the essential role of molecular chirality in realizing such chiral superconductivity. Critical current measurements at low temperature further show asymmetric phase shifts under opposite supercurrent, with the phase difference approaching π at the lowest temperature, further supporting the emergence of topologically non-trivial superconductivity. The ability to rationally design and synthesize artificial chiral materials that can host prominent characteristics of chiral superconductivity could transform the field and offer unprecedented opportunities for tuning topological phases of matter. Our study signifies the potential of hybrid superlattices with intriguing coupling between the crystalline atomic layers and the self-assembled molecular layers. It could chart a versatile path to artificial quantum materials by combining a vast library of layered crystals of rich physical properties with the nearly infinite choices of molecular systems. With versatile molecular design strategies, a vast library of chiral molecules of distinct structural motifs and functional groups can be synthesized and combined with solid-state 2DACs, offering unprecedented versatility in tailoring their topology and exotic physical properties by design.

**Methods**

**Synthesis of MBA/H-TaS$_2$.** H-TaS$_2$ was synthesized by heating a sealed ampoule with sulfur and tantalum inside with different heating procedures. (S)-(−)-α-Methylbenzylamine (98%) (S-MBA) and the-(+)-α-Methylbenzylamine (98%) (R-MBA) were purchased from Sigma-Aldrich. 20 mg 2H-TaS$_2$ was placed in a glass vial with 1 mL S- or R-MBA. The intercalations process was performed simply by stirring the mixture at 65 °C for more than 48 hours under N$_2$ atmosphere.

**Fabrication of chiral intercalated 2H-TaS$_2$ superconducting rings**. The superconducting rings were fabricated using a standard e-beam lithography and reactive-ion etching technique. First, 2H-TaS$_2$ flakes with a typical thickness between 10-20 nm were exfoliated using Nitton tape from bulk single crystals to SiO$_2$/Si substrates. Next, Ti/Au contacting electrodes to the flakes were prepared by electron beam lithography followed by electron beam evaporation. After the metallization and lift-off, etching patterns were defined by electron beam lithography, then etched by reactive ion etching with CF$_4$/O$_2$. Lastly, the devices were immersed in MBA solution and stirred at 65 °C for more than 48 hours under N$_2$ atmosphere, followed by isopropanol cleaning.

**Low temperature transport measurements**. The devices are wire bonded to the chip carrier for electrical and magnetotransport characterizations in a physical properties measurement system with a dilution refrigerator unit (Quantum Design, lnc.). Low temperature transport measurements were carried out using a standard four-probe measurement with the standard AC lock-in technique. The AC excitation current is 100 nA, unless otherwise specified. Each time before cooling down the devices, the superconducting magnet are oscillated to zero field to minimize the flux trapping. To reduce the time-domain fluctuations, the magnetic field is scanned with various speed between 1 Oe/s to 5 Oe/s for different traces. The critical current mapping is performed with a DC setup using Keithley 6221 current sourcemeter and Keithley 2182A nanovoltmeter. The differential resistance is numerically derived from the DC data, the critical current is extracted by finding the DC bias at which the differential resistance exceeds 50% of the normal resistance. To extract the phase shift, external flux values were read off from the minima of $I_C$ oscillations, and the then linearly fitted to the nearest integer. The intercepts of the linear fitting divided by the oscillation period give the relative phase shift.

**Data availability**. The data that supports the plots within this paper and other findings of this study are available from the corresponding author upon reasonable request.

**Acknowledgements.** The authors acknowledge the helpful discussion with Dr. Yuan Ping and Dr. Leonid Rokhinson. The authors would also like to thank Jingyi Fang for the kind support during the measurements.

**Author Contributions** Xiangfeng Duan, Z. W. and G. Q. conceived and designed the research. H. R. developed synthetic methods. Z. W. performed device fabrication with the help from B. Z., D. X. and L.W.. Z. W. and G. Q. performed the electrical measurements and data analysis with the help from Q. Q.. J. Y. Z. and J. X. Z. contributed to discussions and helped analyzed the data. K. L. W. and Xiangfeng Duan supervised the research. Z. W., G. Q., and Xiangfeng Duan co-wrote the manuscript with inputs from all the authors. All authors discussed the results and commented on the manuscript.

**Competing interests.** The authors declare no competing financial interests.

**Additional information**

**Correspondence and requests for materials** should be addressed to and Xiangfeng Duan.

**Reprints and permissions information** is available at www.nature.com/reprints.

**Publisher's note:** Springer Nature remains neutral with regard to jurisdictional claims in published maps and institutional affiliations.

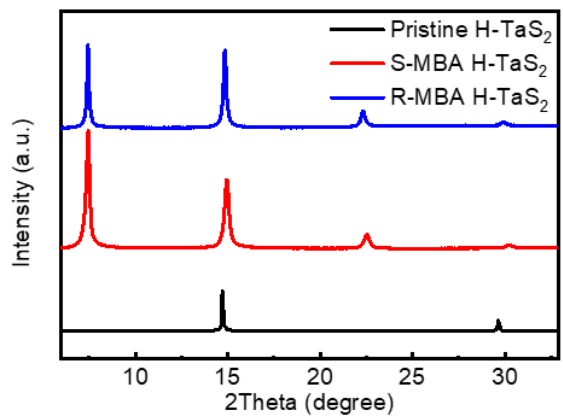

**Extended Data Fig. 1| X-ray diffraction (XRD) patterns of the R-MBA and S-MBA intercalated 2H-TaS₂.** XRD patterns of the chiral R- and S-MBA/H-TaS$_2$ CMISs and pristine H-TaS$_2$. The resulting intercalation superlattices exhibit a notable expansion of interlayer spacing from 5.8 Å to 11.7 Å, with the sharp diffraction peaks highlighting the formation of highly ordered superlattice structures.

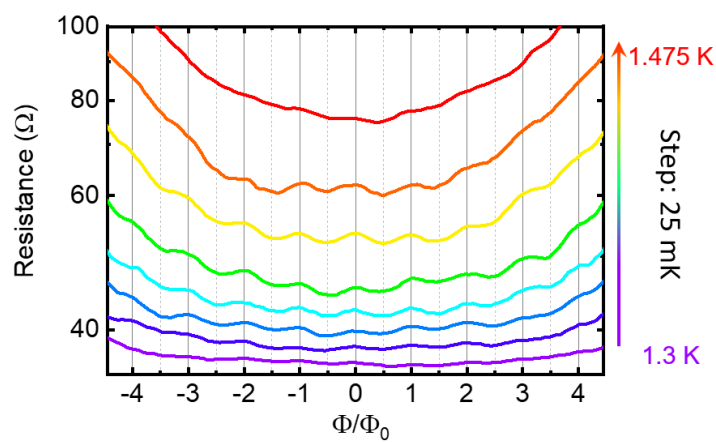

**Extended Data Fig. 2| π-phase shift of S-MBA intercalated device.** Four-terminal resistance (logarithmic scale) of an S-MBA intercalated 2H-TaS$_2$ Little-Parks device as a function of magnetic field at different temperatures. The color map in Fig. 2e was extracted from these data.

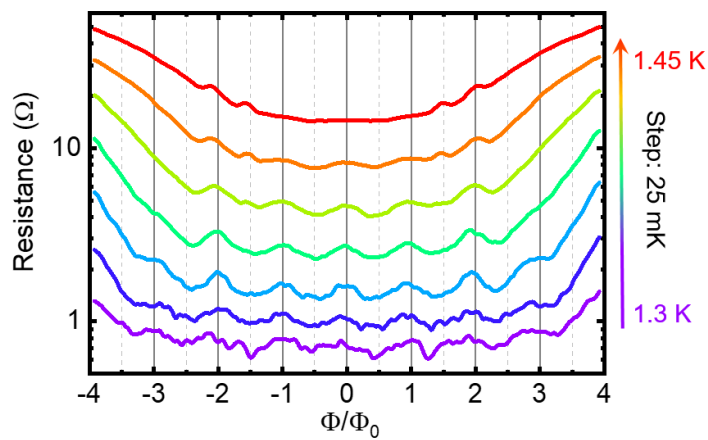

**Extended Data Fig. 3| Robust π-phase shift of chiral molecule intercalated device after thermal cycling.** The Little-Parks effect in the same device as in Fig. 2c and 2d remeasured after being warmed up above $T_C$ and cooled to the base temperature. A robust π-phase shift can be repeated after thermal cycling.

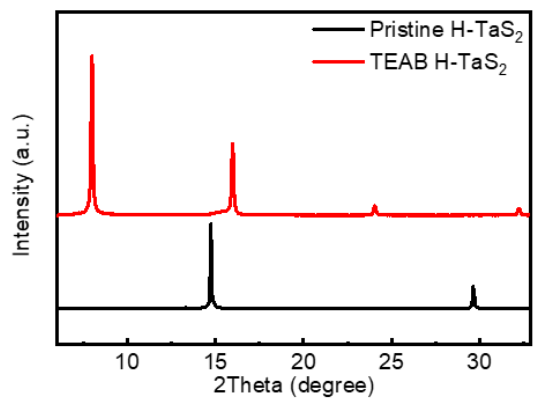

**Extended Data Fig. 4| X-ray diffraction (XRD) pattern of the achiral molecule intercalated 2H-TaS$_2$.** XRD patterns of the achiral TEAB intercalated H-TaS$_2$/T-TaS$_2$ CMISs and pristine H-TaS$_2$ showing sharp diffraction peaks with a significant expansion of interlayer spacing from 5.8 Å to 11.1 Å, highlighting the formation of highly ordered superlattice structures.

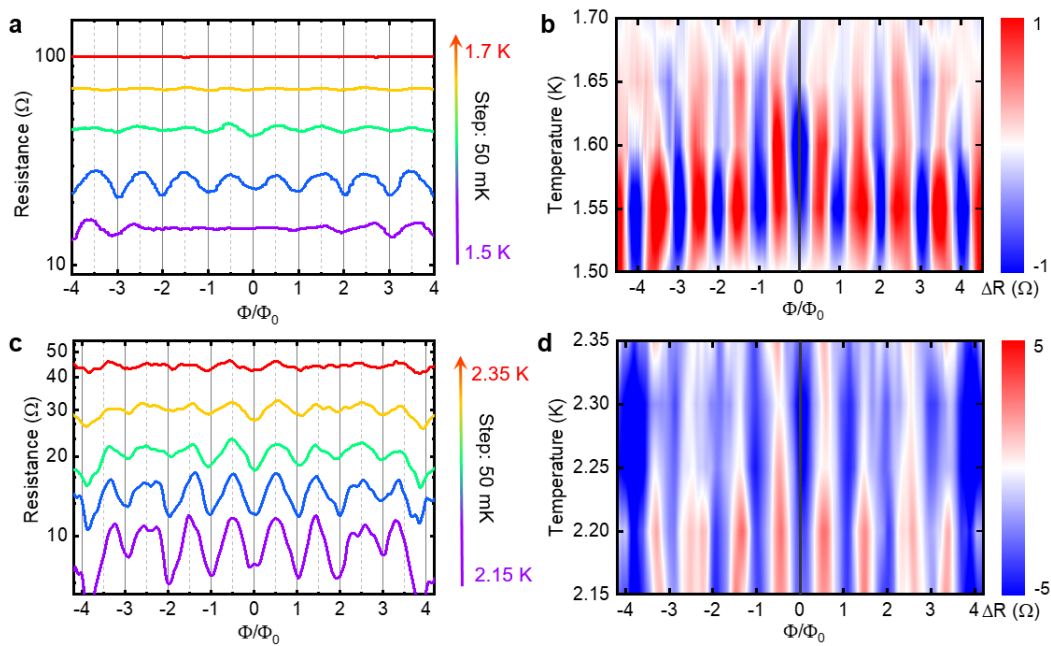

**Extended Data Fig. 5| Additional achiral molecule intercalated 2H-TaS$_2$ devices. a, c**, Four-terminal resistance (logarithmic scale) of two additional achiral intercalated 2H-TaS$_2$ Little-Parks devices as a function of magnetic field at different temperatures. **b** and **d**, Color mapping of the Little-Parks oscillation amplitudes of chiral molecule intercalated devices as a function of the temperature and the external magnetic flux. No phase shift of the superconducting ordering parameter was observed in the achiral molecule intercalated 2H-TaS$_2$ superconducting ring, highlighting the essential role of molecular chirality in achieving π-phase shift.

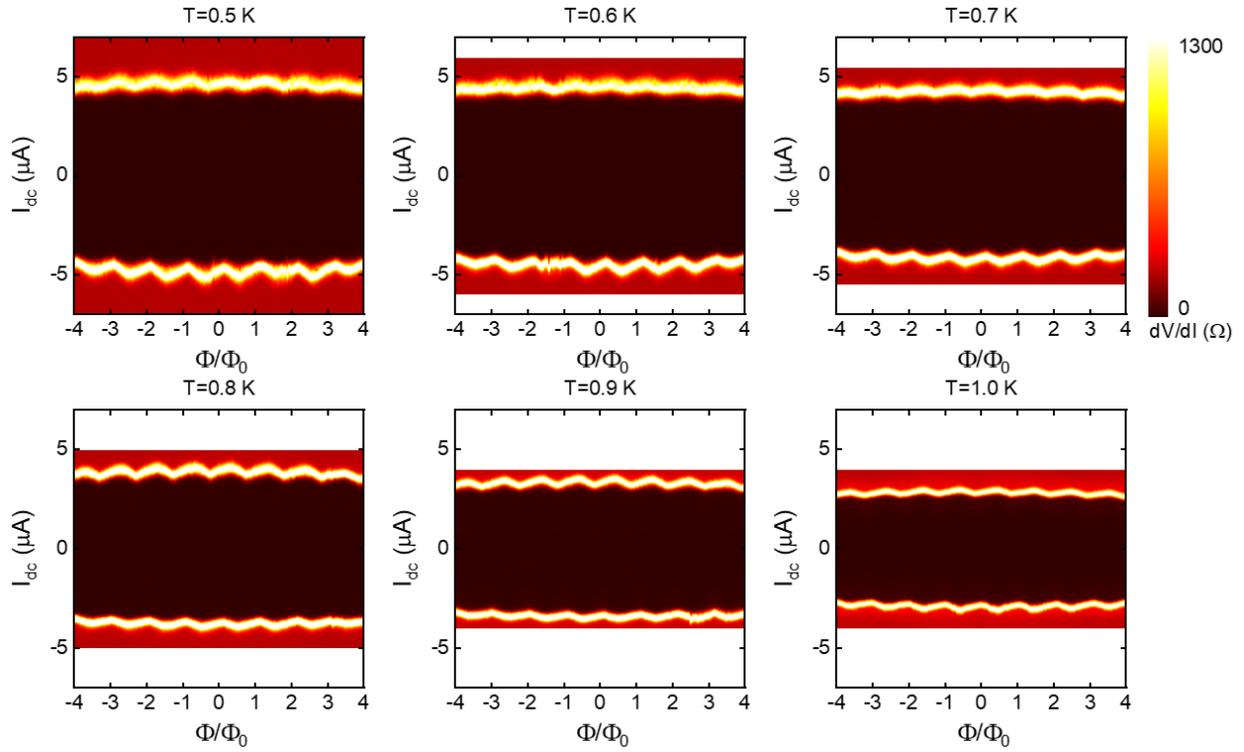

**Extended Data Fig. 6|** Additional color mapping of the differential resistance of the R-MBA intercalated $2H\text{-}TaS_2$ ring as a function of the temperature and the external magnetic flux.